\documentstyle[12pt]{article}
\begin{document}
\baselineskip=24pt
\begin{center}
{\bf Temperature Shift of Neutrino Energy in
the External Magnetic Field}
\end{center}
\vspace{0.5 cm}
\centerline {M.~Gogberashvili, L.~Midodashvili, P.~Midodashvili}
\centerline {\sl Institute of Physics, Georgian Academy of Sciences, 380077
Tbilisi, Georgia}      
\centerline {\sl e-mail: gogber@physics.iberiapac.ge}
\vspace{0.5 cm}
\begin{abstract}
It is evaluated the first order temperature correction to the energy of 
massive neutrino in the case of weak magnetic field
and temperatures $T \ll M_{W}$.
\end{abstract}
\vspace{0.5 cm}
~~~~~~~~~~moment.\\ 
\vspace{0.5 cm}
{\bf PACS indeces:} 13.10.+q; 13.40.-f
\vspace{1 cm}

~~~~~Finite temperature field theories have often been discussed in physics
for various reasons \cite{1, 2, 3, 4}. The most fundamental reason is
that physics experiments are not carried out at zero temperature;
however, since relativistic field theories usually deal with energy
scales characterized, as a minimum, by the electron mass, laboratory
ambient temperatures qualify as low, and temperature corrections are
justifiably neglected. In recent years theoretical interest in
finite temperature field theory has grown for other reasons. One is
the increased interest in nonabelian gauge theories with spontaneous
symmetry breaking; at zero temperature they exhibit a broken-symmetry
phase, but at high temperatures the symmetry commonly is restored.
This leads into cosmological interests; at the epochs when
elementary-particle phenomena were relevant to the evolution of our
universe, the temperature was also extremely high, and a satisfactory
relation between cosmological evidence and particle physics must take into
account temperature effects. A separate interest is related to the
increasing availability of data on high energy hadronic collisions; a
common feature of models of such processes is the assumption of a temporary
state of high-temperature equilibrium before the final state forms. To
improve such models a better understanding of finite-temperature field
theory will be needed. It is also worth noticing that recently there has
been a great interest in the problem of neutrino interactions with a
thermal background \cite{11, 12, 13, 14}.

In this paper we present a calculation of finite temperature corrections to
massive neutrino energy in the external magnetic field. 
Our fundamental definition of "finite temperature" is the following: "At finite
temperature" means "in the presence of particles". In this case in a
thermal-equilibrium distribution exist such particles as photons,
electrons, positrons and intermediate bosons. Calculations are carried out
in the frame of Weinberg-Salam theory of electroweak interactions
\cite{5, 7, 8}.

The part of Lagrangian which is connected with dynamic nature of
anomalous magnetic moment of Dirac massive neutrino is
\begin{equation}
L = L_{W} + L_{\nu} + L_{e} + L_{int} + L_{g} ,
\label{1}
\end{equation}
where
\begin{equation}
L_{W} = - \frac{1}{2}[D_{\mu}W^{+}_{\nu} - D_{\nu}W^{+}_{\mu}]^{2} + \
M^{2}W^{+}_{\mu}W^{-\mu} - ie(\partial_{\mu}A_{\nu} - \
\partial_{\nu}A_{\mu})W^{+\mu}W^{-\nu}  \nonumber
\label{2}
\end{equation}
is the Lagrangian of $W$ - boson field interacting with electromagnetic
field,
\begin{equation}
L_{\nu} = \bar{\nu}(i\gamma^{\mu}\partial_{\mu} - m_{\nu})\nu
\label{3}
\end{equation}
is the Lagrangian of free neutrino,
\begin{equation}
L_{e} = \bar{e}(i\gamma^{\mu}\partial_{\mu} - e\gamma^{\mu}A_{\mu} - m)e
\label{4}
\end{equation}
is the Lagrangian of electron field in the external magnetic field
$A_{\mu}$,
\begin{equation}
L_{int} = \frac{g}{\sqrt{2}}(\bar{\nu_{L}}W^{+}_{\mu}\gamma^{\mu}e_{L} +
\bar{e_{L}}W^{-}_{\mu}\gamma^{\mu}\nu_{L})
\label{5}
\end{equation}
is the Lagrangian of interaction of electron, neutrino and $W$-boson,
\begin{equation}
L_{g} = -
\frac{1}{\xi}(D^{+\mu}W^{-}_{\mu})(D^{-\alpha}W^{+}_{\alpha})
\label{6}
\end{equation}
is the part fixing gauge.

In the equations (\ref{2}) and (\ref{6}) $D^{\pm}_{\mu} =
\partial_{\mu} \pm ieA_{\mu}$, and right and left components of Dirac
bespinors are equal to $
\psi_{L} = \frac{1}{2}(1 + \gamma^{5})\psi ,~~
\psi_{R} = \frac{1}{2}(1 - \gamma^{5})\psi .
$

It is known that the contribution of charged scalars $\varphi$ to radiative
energy shift of neutrino is smaller than $W$ - boson contribution:
\begin{equation}
\frac{\Delta E_{\varphi}}{\Delta E_{W}} \sim
\frac{m_{e}^{2}}{M_{W}^{2}} \ll 1 ,
\label{8}
\end{equation}
therefore the corresponding part of the Lagrangian (\ref{1}) is
omitted and we consider only $W$ - boson contribution to the
radiative shift of neutrino energy. Let us consider electron neutrino
with four-momentum $q = (q_{0}, \vec{q})$ in the heat "bath" of
electrons, positrons, photons and intermediate bosons with zero
chemical potential in the external magnetic field
\begin{equation}
A^{\mu} = (0, 0, xH, 0) .
\label{9}
\end{equation}

The selftime representation of $W$ - boson field propagator in
Feynman gauge is given by
\begin{equation}
B_{\mu\alpha}(x, x') =
\frac{1}{(4\pi)^{2}}\int_{0}^{+\infty}\frac{dt}{t^{2}} \
B_{\mu\alpha}(t)exp(-iM^{2}t)\frac{bt}{\sin (bt)} \
exp[-\frac{i}{4}(x^{2}_{\parallel} - \frac{bx^{2}_{\perp}}{\tan
(bt)} - 4\Omega)] , \label{10} \end{equation} where \begin{eqnarray}
\Omega = -b(x_{2} - x_{2}')(x_{1} + x_{1}'), \\ \nonumber
x^{2}_{\perp} = (x_{1} - x_{1}')^{2} + (x_{2} - x_{2}')^{2} , \\ b =
eH,~~~~~ e > 0 ,~~~~ \nonumber \label{11} \end{eqnarray}
and matrix $B(t)$ has the following nonzero elements:
\begin{eqnarray} B_{00} =
- B_{33} = 1 , ~~~~~B_{22} = B_{11} = - \cos (2y) , \\ \nonumber B_{12} =
- B_{21} = -\sin (2y) ,~~~~~ y = bt .  \label{12} \end{eqnarray}

The finite temperature electron propagator in the real-time
representation is given by
\begin{equation}
G_{\beta}(x, x') = -
\frac{1}{2\pi}\int_{0}^{+\infty}d\omega e^{i\omega(t - t')}\sum_{s}
\frac{\psi^{(\varepsilon)}_{n}(\vec{x})\bar{\psi}^{(\varepsilon)}_{n}(\vec{x}')}
{\omega + \varepsilon E_{n}(1 - i\delta)} 
+ i\sum_{s}e^{-i\varepsilon E_{n}(t - 
t')}\frac{\varepsilon\psi^{(\varepsilon)}_{n}(\vec{x})
\bar{\psi}^{(\varepsilon)}_{n}(\vec{x}')}{1 + exp(E_{n}/T)} ,
\label{13}
\end{equation}
where $\psi^{(\varepsilon)}_{n}(\vec{x})$ is the solution of Dirac
equation in the field (\ref{9}) \cite{9}:
\begin{equation}
\psi^{(\varepsilon)}_{n}(x) = e^{-i\varepsilon E_{n}t + ip_{2}y +
ip_{3}z}N\left( \begin{array}{c}
C_{1}u_{n - 1}(\eta) \\
iC_{2}u_{n}(\eta) \\
C_{3}u_{n - 1}(\eta) \\
iC_{4}u_{n}(\eta)
\end{array} \right)
\label{14}
\end{equation}

In this expression $u_{n}(\eta)$ is Hermite function with argument
\begin{equation}
\eta = \sqrt{eH}(x_{1} - \frac{p_{2}}{\sqrt{eH}}) ,
\label{15}
\end{equation}
$C_{i}~~~ (i = 0,1,2,3)$ are spin coefficients \cite{9}, and
\begin{equation}
N = (eH)^{1/4}/L
\label{16}
\end{equation}
is normalization factor with normalization length equal to $L$.

The shift of neutrino energy is given by
\begin{equation}
\Delta E_{\nu}(H,T) = - \frac{ig^{2}}{2} \int\!\!\int
^{+\infty}_{-\infty}d^{4}xd^{4}x' \bar{\nu}_{L}(x)
\gamma^{\mu}G_{\beta}(x,x')\gamma^{\delta}\nu_{L}(x')B_{\mu\delta}(x,x') .
\label{17}
\end{equation}

Let us consider pure temperature dependence part of expression
(\ref{17}). In the real time formalism \cite{1, 2, 3, 4}  one gets
automatically the temperature dependence separated from the
zero-temperature terms.

If we consider neutrino moving along $x$ -
axis, than neutrino wave-function has form:
\begin{equation}
\nu(x) = L^{-3/2}e^{-iE_{\nu}t + iq_{1}x}b_{\lambda} .
\label{18}
\end{equation}

In (\ref{18}) we must take $\lambda = 1$ when
the neutrino spin is orientated along the magnetic field, and
\begin{equation}
b_{1} = \left( \begin{array}{c}
q_{1}/\sqrt{2E_{\nu}(E_{\nu} - m_{\nu})} \\ 0 \\ 0 \\ \sqrt{(E_{\nu}
- m_{\nu})/2E_{\nu}} \end{array} \right) =
\left( \begin{array}{c} A \\ 0 \\ 0 \\ B \end{array} \right) .
\label{19}
\end{equation}

 In the case when the neutrino spin is orientated against the
 magnetic field we must take $\lambda = -1$ , and
\begin{equation}
b_{-1} = \left( \begin{array}{c}
0 \\ q_{1}/\sqrt{2E_{\nu}(E_{\nu} - m_{\nu})} \\
\sqrt{(E_{\nu} - m_{\nu})/2E_{\nu}} \\ 0 \end{array} \right) = \left(
\begin{array}{c} 0 \\ A \\ B \\0 \end{array} \right) .
\label{20} \end{equation}

In the expressions (\ref{19}) and (\ref{20}) the neutrino energy
is equal to $E_{\nu} = \sqrt{m_{\nu}^{2} + q^{2}_{1}}$. Putting
expressions of $G_{\beta}, B_{\mu\alpha}$ end $\nu_{L}$ in the
equation (\ref{17}) and taking the sum over the spins of
intermediate electron states, we get the following formula for
temperature shift of neutrino energy:
\begin{eqnarray}
\Delta E_{\nu}(H,T) =
\frac{ig^{2}}{32\pi^{2}L^{7}}\int\!\!\int^{+\infty}_{-\infty}d^{4}xd^{4}x' \
e^{iE_{\nu}(x_{0} - x_{0}') - iq_{1}(x_{1} -x_{1}')} \times \\ \nonumber
\sum_{n, \varepsilon, p_{2}, p_{3}}\frac{\varepsilon}{1 +
exp(E_{n}/T)}e^{i\varepsilon E_{n}(x_{0}' - x_{0}) + ip_{2}(x_{2} -
x_{2}') + ip_{3}(x_{3} - x_{3}')}\int^{+\infty}_{0}\frac{dt}{\cos (eHt)}
e^{\frac{i}{2}eH(x_{1} + x_{1}')}\times \\ \nonumber
\int^{+\infty}_{-\infty}d^{4}k exp[-ik(x - x') + it(k_{0}^{2} - k^{2}_{3}
- \frac{\tan (eHt)}{eHt}k^{2}_{\perp} - M^{2})]F(\lambda, n,
\varepsilon, p_{3}, \eta, \eta') , \label{21} \end{eqnarray}
where the sums are taking over the main quantum number $n$, over the
momenta $p_{2}, p_{3}$ and over the sign coefficient $\varepsilon =
\pm1$. In (\ref{20}) the function $F$ has the following form:
\begin{eqnarray} F = -i\frac{2\varepsilon\sqrt{4\gamma n}}{E_{n}} AB[
u_{n - 1}(\eta)u_{n}(\eta') -
u_{n}(\eta)u_{n - 1}(\eta')] - \\ \nonumber - (1 +
\varepsilon\frac{p_{3}}{E_{n}})e^{i2eHt}[A^{2}(1 +
\lambda) + B^{2}(1 - \lambda)]u_{n}(\eta)u_{n}(\eta') - \\ - (1 -
\varepsilon\frac{p_{3}}{E_{n}}) e^{-i2eHt}[A^{2}(1 - \lambda) +
B^{2}(1 + \lambda)]u_{n - 1}(\eta)u_{n - 1}(\eta')  .  \nonumber
\label{22}
\end{eqnarray}

Integrating over the variables $x_{2}, x_{2}', x_{3}, x_{3}'$ and
over the $x_{1}'$ (making substitution $x_{1} \rightarrow y =
\frac{\sqrt{eH}}{2}(x_{1} - x_{1}'),~~~ x_{1}' \rightarrow
x_{1}'$), we get expression
\begin{eqnarray} \Delta E_{\nu}(H,T) =
\frac{ig^{2}}{32\pi^{2}\sqrt{eH}}\sum_{n,\varepsilon}
\int^{+\infty}_{-\infty}dy\int^{+\infty}_{o}dt
\int^{+\infty}_{-\infty}\frac{d^{4}k}{\cos (eHt)}
[1 +exp(E_{n}/T)]^{-1}\times \\ \nonumber exp\{it[(E_{\nu} - \varepsilon
E_{n})^{2} - k_{3}^{2} - \frac{\tan (eHt)}{eHt}
k^{2}_{\perp} - M^{2}] + i\frac{2(k_{1} - q_{1})y}{\sqrt{eH}}\}F ,
\label{23} \end{eqnarray}
where $F$ is the same as in the formula (\ref{21}), but with the 
following variables:  
\begin{equation} \eta = y - \frac{k_{2}}{\sqrt{eH}} ,~~~~~ \eta' 
= -y - \frac{k_{2}}{\sqrt{eH}} ,~~~~~p_{3} = -k_{3} .  \label{24}
\end{equation}

Using table formula \cite{10}:
\begin{eqnarray}
\int^{+\infty}_{-\infty}e^{-x^{2}}H_{m}(x + y)H_{n}(x + z)dx =
2^{n}\pi^{1/2}m!z^{n - m}L^{n - m}_{m}(-2yz) ;~~~~~ n \geq m ,
\label{25}
\end{eqnarray}
(here $H_{n}(x)$ is Hermite polynomial and $L^{n - m}_{m}(x)$ -
Laguerre polynomial), integrating over $y$  and making substitution of 
variebles
\begin{eqnarray}
\frac{k_{1} - q_{1}}{\sqrt{eH}} = x_{1},~~~~~
\frac{k_{2}}{\sqrt{eH}} = x_{2}, \\ \nonumber
\frac{k_{3}}{\sqrt{eH}} = x_{3},~~~~~eHt = y  .
\label{26}
\end{eqnarray}
we get
\begin{eqnarray}
\Delta E_{\nu}(H,T) = \frac{ig^{2}\sqrt{eH}}{32\pi^{2}}\sum_{n,\varepsilon}
\int^{+\infty}_{0}\frac{dy}{\cos y}
\int^{+\infty}_{-\infty}\frac{(-1){n}\varepsilon d^{3}x}{1 + exp(E_{n}/T)}
\times \\ \nonumber
exp\{iy[\frac{(E_{\nu} - \varepsilon E_{n})^{2}}{eH} - x_{3}^{2} -
\frac{\tan y}{y}((x_{1} + \frac{q_{1}}{\sqrt{eH}})^{2} + x^{2}_{2})
 - \frac{M^{2}}{eH}]\}\times \\ \nonumber
exp[- iy(x_{1}^{2} + x_{2}^{2})][i2x_{1}R_{n - 1,n}
L^{1}_{n - 1}(2x^{2}_{1} + 2x^{2}_{2}) + \\
+ R_{n,n}L^{0}_{n}(2x^{2}_{1} + 2x^{2}_{2}) - R_{n - 1, n - 1}L^{0}_{n - 1}
(2x^{2}_{1} + 2x^{2}_{2})] , \nonumber
\label{29}
\end{eqnarray}
where
\begin{eqnarray}
R_{n - 1, n} = - \frac{i2\varepsilon\sqrt{4\gamma n}AB}{E_{n}} ,~~~~~~~ 
~~~~~~~~~~~~~~~\\ \nonumber
R_{n - 1, n - 1} = - (1 - \varepsilon\frac{k_{3}}{E_{n}})
[A^{2}(1 - \lambda) + B^{2}(1 + \lambda)]exp(- i2eHt) , \\
R_{n, n} = - (1 + \varepsilon\frac{k_{3}}{E_{n}})
[A^{2}(1 + \lambda) + B^{2}(1 - \lambda)]exp(i2eHt) , \nonumber
\label{27}
\end{eqnarray}

In the plane of variables $x_{1}$ and $x_{2}$ we can pass to polar
coordinates $x$ and $\varphi$ , where
\begin{eqnarray}
x_{1} = x\cos \varphi,~~~~~ x_{2} = x\sin \varphi, \\ \nonumber
\int^{+\infty}_{-\infty}dx_{1}dx_{2} \Rightarrow
\int^{+\infty}_{0}xdx\int^{2\pi}_{0}xd\varphi .
\label{30}
\end{eqnarray}

After such substitution we can use well known table formulae \cite{10}:
\begin{eqnarray}
\int^{2\pi}_{0}d\varphi e^{-ix\sin \varphi + in\varphi} = 2\pi
J_{n}(x); ~~~~~~~~~~~~~~~~~~~~~~~~~~~~~~~~~~\\ \nonumber
\int^{\infty}_{0}x^{\nu + 1}e^{-\beta x^{2}}L^{\nu}_{n}(\alpha x^{2})
J_{\nu}(xy)dx = 2^{-\nu - 1}\beta^{-\nu - n - 1}(\beta - \alpha)^{n}
y^{\nu}e^{-y^{2}/4\beta}L^{\nu}_{n}(\frac{\alpha y^{2}}{4\beta(
\alpha - \beta)}) . 
\label{31}
\end{eqnarray}
($J_{n}(x)$ is Bessel function) and integrate over $\varphi$ and $x$ .

Finally for the temperature shift of neutrino energy we obtain the
following expression
\begin{equation}
\Delta E_{\nu}(H,T) = \Delta E_{1}(H,T) + \Delta E_{2}(H,T) ,
\label{32}
\end{equation}
where
\begin{eqnarray}
\Delta E_{1}(H,T) = \frac{ig^{2}\sqrt{eH}}{16\pi^{2}}\sum_{n,\varepsilon}
\int^{+\infty}_{0}dy
\int^{+\infty}_{-\infty}\varepsilon dx[1 + exp(E_{n}/T)]^{-1}\times \\
\nonumber
exp[iy\frac{m^{2}_{\nu} + m^{2} + q^{2}_{1} - 2\varepsilon E_{\nu}E_{n}
- M^{2}}{eH} -  i\sin y e^{-iy}\frac{q^{2}_{1}}{eH}]\times \\ 
\{i2\varepsilon\frac{q^{2}_{1}}{E_{\nu}E_{n}} \sin y L^{1}_{n - 1}(z)
- \frac{1}{2}[e^{iy}L^{0}_{n}(z) + e^{-iy}L^{0}_{n - 1}(z)]\}
\nonumber \label{33} \end{eqnarray}
is part which does not depend on neutrino spin orientation, and 
\begin{eqnarray} \Delta E_{2}(H,T) =
\frac{ig^{2}\sqrt{eH}}{16\pi^{2}}\sum_{n,\varepsilon}
\int^{+\infty}_{0}dy
\int^{+\infty}_{-\infty}\varepsilon dx[1 + exp(E_{n}/T)]^{-1}\times \\
\nonumber
exp[iy\frac{m^{2}_{\nu} + m^{2} + q^{2}_{1} - 2\varepsilon E_{\nu}E_{n}
}{eH} -  i\sin y e^{-iy}\frac{q^{2}_{1}}{eH}]
(- \frac{\lambda}{2})\frac{m_{\nu}}{E_{\nu}}
[e^{iy}L^{0}_{n}(z) - e^{-iy}L^{0}_{n -
1}(z)] \nonumber \label{34} \end{eqnarray}
is the part which depends on neutrino spin orientation.

Below we present results for various values of temperature and neutrino
energies:

1. In the case of weak magnetic fields $H \ll m^{2}/e$  and low
temperatures $T/m \ll 1,~~~ \Delta E_{1}(H,T)$ and $\Delta
E_{2}(H,T)$ are of order $exp(-m/T)$; i.e.  we have exponential
suppression.

2. In the case of weak magnetic fields $H \ll m^{2}/e$  and
temperatures $m\ll T\ll M$ for neutrino energies
$E_{\nu}\ll M^{2}/T$ we have
\begin{eqnarray}
\Delta E_{1}(H,T) = \frac{g^{2}}{4\pi^{2}}\frac{7\pi^{4}}{60}E_{\nu}
(\frac{T}{M})^{4}, \\ \nonumber
\Delta E_{2}(H,T) = \lambda
(\mu_{0}H)\frac{4\pi^{2}}{9}(\frac{T}{M})^{2} .
\label{35}
\end{eqnarray}
In this formulae $\mu_{0} = 3g^{2}m_{\nu}e/64\pi^{2}M^{2}$ is the
statical neutrino magnetic moment.

\vspace{0.3cm}
We would like to thank V. Ch. Zhukovsky, P. A. Eminov, I. M. Ternov and 
other members of stuff of Moscow University Theoretical Physics Department 
for usefully discussions and to N. Abuashvili for helping. This work was 
supported in part by International Science Foundation (ISF) under the Grant 
No MXL000.

\end{document}